\documentclass{article}
\usepackage{spconf,amsmath,graphicx}
\usepackage{blindtext}
\usepackage{multirow}
\usepackage{cleveref}

\usepackage{enumitem}
\usepackage{etoolbox}
\usepackage{graphicx}
\usepackage{makecell}
\usepackage{caption}
\usepackage{wrapfig}
\usepackage{setspace}
\usepackage{subcaption}
\usepackage{floatflt}
\usepackage{float}
\usepackage{setspace}

\renewrobustcmd{\bfseries}{\fontseries{b}\selectfont}
\renewrobustcmd{\boldmath}{}
\newrobustcmd{\B}{\bfseries}

\newsavebox\CBox

\makeatletter

\renewcommand{\section}{\@startsection
   {section}%
   {1}%
   {}%
   {-0.4\baselineskip}%
   {0.2\baselineskip}%
   {}}%

\renewcommand{\subsection}{\@startsection
  {subsection}%
  {2}%
  {}%
  {-0.1\baselineskip}%
  {0.1\baselineskip}%
  {}}%

\renewcommand{\subsubsection}{\@startsection
  {subsubsection}%
  {3}%
  {}%
  {-0.2\baselineskip}%
  {0.2\baselineskip}%
  {}}%

\Crefname{equation}{Eq.}{Eqs.}
\Crefname{figure}{Fig.}{Figs.}
\Crefname{tabular}{Tab.}{Tabs.}
\Crefname{table}{Tab.}{Tabs.}
\Crefname{section}{Sec.}{Sec.}

\DeclareMathOperator*{\argmax}{arg\,max}

\title{Lattice-Free Sequence Discriminative Training for Phoneme-Based Neural Transducers}
\name{Zijian Yang$^1$, Wei Zhou$^{1,2}$, Ralf Schlüter$^{1,2}$, Hermann Ney$^{1,2}$}
\address{$^1$Human Language Technology and Pattern Recognition, Computer Science Department,\\
RWTH Aachen University, 52074 Aachen, Germany,\\
$^2$AppTek GmbH, 52062 Aachen, Germany}
\begin{document}
%
\maketitle
\begin{abstract}
Recently, RNN-Transducers have achieved remarkable results on various automatic speech recognition tasks. However, lattice-free sequence discriminative training methods, which obtain superior performance in hybrid models, are rarely investigated in RNN-Transducers. In this work, we propose three lattice-free training objectives, namely lattice-free maximum mutual information, lattice-free segment-level minimum Bayes risk, and lattice-free minimum Bayes risk, which are used for the final posterior output of the phoneme-based neural transducer with a limited context dependency. Compared to criteria using N-best lists, lattice-free methods eliminate the decoding step for hypotheses generation during training, which leads to more efficient training. Experimental results show that lattice-free methods gain up to 6.5\% relative improvement in word error rate compared to a sequence-level cross-entropy trained model. Compared to the N-best-list based minimum Bayes risk objectives, lattice-free methods gain 40\% - 70\% relative training time speedup with a small degradation in performance.
\end{abstract}
\section{Introduction \& Related Work}
\label{sec:intro}

Nowadays, sequence-to-sequence (seq2seq) modeling methods have gained great success in automatic speech recognition (ASR) tasks. Various modeling approaches like attention-based encoder-decoder (AED) models \cite{bahdanau2016end, chan2016listen, Tske2020SingleHA}, connectionist temporal classification (CTC) \cite{graves2006connectionist}, and recurrent neural network transducer (RNN-T) \cite{graves2012sequence} are proposed. Among these approaches, RNN-T receives a huge interest because it is suitable for streaming tasks with competitive performance \cite{rao2017exploring}.

Sequence discriminative training criteria have been shown to improve ASR models \cite{vesely2013sequence, Shannon2017OptimizingEW}. Most popular criteria include maximum mutual information (MMI) \cite{bahl1986maximum}, boosted MMI (bMMI) \cite{povey2008boosted}, minimum phone error (MPE) \cite{vesely2013sequence,povey2005discriminative}, minimum word error rate (MWE) \cite{vesely2013sequence,povey2005discriminative,Guo2020EfficientMW, prabhavalkar2018minimum} and state-level minimum Bayes risk (sMBR) \cite{vesely2013sequence, senior2015acoustic}. 
These methods usually require on-the-fly decoding to generate lattices or N-best lists for hypotheses space of discrimination, which is time and resource-wise costly. 
To make the training more efficient, \cite{povey2016purely} proposed lattice-free MMI (LF-MMI) for hybrid models, which spares this decoding step. 
Later on, other LF methods like LF-sMBR \cite{kanda2018lattice, Michel2019ComparisonOL} and LF-bMMI \cite{zhang2021lattice} were introduced for hybrid/CTC models. 
LF-MMI was also applied to AED and RNN-T models \cite{tian2022consistent, tian2022integrating} as an auxiliary loss on the encoder output, rather than on the final posterior output.
In general, the full context dependency of such seq2seq models makes it difficult to directly apply LF methods on the final posterior output.
Recently, \cite{zhou2022efficient, zhou2021phoneme} showed that phoneme-based neural transducer with limited context dependency can also achieve superior performance, which allows to directly apply these LF methods on the output of transducer models.

In this paper, we propose three kinds of LF training objective functions for phoneme-based neural transducers. Compared to \cite{tian2022consistent, tian2022integrating}, we use these objective functions for the final posterior output of neural transducers with limited context dependency.
Compared to criteria using N-best lists, our methods avoid decoding during training and thus, make the training more efficient. 
Experiments on Librispeech \cite{panayotov2015librispeech} show that our proposed criteria give competitive improvements over the baseline as N-best list based MBR, but with a significant training speedup.
Besides applying LF training criteria upon the baseline transducer model, we also explore replacing the sequence-level cross-entropy (CE) criterion with LF-MMI, which can be difficult for N-best list based methods due to the possible poor quality of the generated N-best list.
Experimental results show that in this case, the model can converge with fewer epochs and obtain a slightly better performance.



\section{Phoneme-based transducer}
In this work, we employ the strictly monotonic RNN-T \cite{Tripathi19} that enforces strictly monotonic alignments between input and output sequences. Given the input sequence $X$, the posterior probability of output label sequence $a_1^S$ is defined as:\\
\scalebox{0.9}{\parbox{1.1\linewidth}{%
\begin{align*}
    \begin{split}
        P_\text{RNNT}(a_1^S|X) &= \sum_{y_1^T: \mathcal{B}(y_1^T)=a_1^S} P_\text{RNNT}(y_1^T|h_1^T)\\
        &= \sum_{y_1^T: \mathcal{B}(y_1^T)=a_1^S} \prod_{t=1}^T P_\text{RNNT}(y_t|\mathcal{B}(y_1^{t-1}), h_t)
    \end{split}
\end{align*}}}
Here $h_1^T$ is the encoder output sequence and $y_1^T$ is the blank $\epsilon$-augmented alignment sequence where $y_t \in \{\epsilon\} \cup \mathcal{V}$ with vocabulary $\mathcal{V}$. $\mathcal{B}$ is the collapse function which maps $y_1^T$ to label sequence $a_1^S$ by removing $\epsilon$ in $y_1^T$. As shown in \cite{ zhou2021phoneme, Prabhavalkar2021LessIM, Ghodis20}, a limited context-dependency can be introduced to simplify the model.\\
\scalebox{0.95}{\parbox{1.05\linewidth}{%
\begin{equation*}
\notag
    P_\text{RNNT}(y_t|\mathcal{B}(y_1^{t-1}), h_t) = P_\text{RNNT}(y_t|a_{s_{t-1}-k+1}^{s_{t-1}}, h_t)
\end{equation*}}}
Here $k$ is the context size and $s_1^T$ is the position sequence with $0 \leq s_t \leq S$ indicating the position in $a_1^S$ where $y_t$ reaches.
Given the target label sequence $\hat{a}_1^{\hat{S}}$, the transducer model can be trained with the sequence-level CE loss by computing the full-sum (FS) over all alignments of $\hat{a}_1^{\hat{S}}$:\\
\scalebox{0.95}{\parbox{1.05\linewidth}{%
\begin{equation*}
\notag
    \mathcal{L}_\text{CE-FS} = - \log P_\text{RNNT}(\hat{a}_1^{\hat{S}}|X)
\end{equation*}}}
For decoding, the decision rule can be formulated as:\\
\scalebox{0.95}{\parbox{1.05\linewidth}{%
\begin{equation*}
\notag
   X \rightarrow \mathcal{W}({a^*}_1^{S^*}) = \argmax_{S, a_1^S} \big[ P_\text{RNNT}(a_1^S|X) \cdot \frac{P^{\lambda_1}_\text{LM}(\mathcal{W}(a_1^S))}{P^{\lambda_2}_\text{ILM}(a_1^S)} \big]
\end{equation*}}}
Here $\mathcal{W}$ is a mapping function that maps the output labels of RNN-T to a word sequence. $P_\text{LM}$ is an external language model (LM) with scale $\lambda_1$ and $P_\text{ILM}$ is the internal language model (ILM) extracted from the RNN-T model with scale $\lambda_2$. In this work, we use zero-encoder \cite{meng2021internal, variani2020hybrid} to extract the ILM.

\section{Lattice-Free Training Objectives}
\label{sec:methodology}
In this section, we discuss three kinds of lattice-free sequence discriminative training objectives: LF-MMI, segment-based MBR (LF-SegMBR), and label-based MBR (LF-MBR). In training, we employ a phoneme-level LM integrated with the RNN-T model. Rather than generating a numerator/denominator graph as in \cite{povey2016purely}, we directly compute the summation by dynamic programming (DP), which will be explained in detail in the following discussion, and we leave the derivative computation to automatic differentiation.




\subsection{MMI}
The MMI training objective is formulated as:\\
\scalebox{0.9}{\parbox{1.1\linewidth}{%
\begin{equation}
    \mathcal{L}_\text{MMI} = -\log \frac{q_\text{seq}(\hat{a}_1^{\hat{S}}|X)}{\sum_{S',{a'}_1^{S'}}q_\text{seq}({a'}_1^{S'}|X)}
    \label{eq:MMI}
\end{equation}}}
where $q_\text{seq}(a_1^S|X)$ is defined as:\\
\scalebox{0.9}{\parbox{1.1\linewidth}{%
\begin{equation*}
    q_\text{seq}(a_1^S|X) = P^\alpha_\text{RNNT}(a_1^S|X) \cdot P^\beta_\text{LM}(a_1^S)
\end{equation*}}}
The numerator in \Cref{eq:MMI} can be computed via the standard RNN-T CE-FS. When the context size is limited to $k$, the recombination for the same limited history $u_1^{k} \in \mathcal{V}^k$ is possible. Therefore, the summation over all sequences in the denominator can be computed by DP:\\
\scalebox{0.9}{\parbox{1.1\linewidth}{%
\begin{equation*}
    \sum_{S',{a'}_1^{S'}}q_\text{seq}({a'}_1^{S'}|X) = \sum_{u_1^k} Q_\text{MMI}(T, u_1^k)
\end{equation*}}}
where the auxiliary function $Q_\text{MMI}$ is defined as:\\
\scalebox{0.9}{\parbox{1.1\linewidth}{%
\begin{equation}
Q_\text{MMI}(t,u_1^k) = \sum_{s,a_1^s:a_{s-k+1}^s=u_1^k} q_\text{seq}(a_1^s|X,t)
\label{eq:mmiauxi}
\end{equation}}}
Here $q_\text{seq}(a_1^s|X,t)$ is the probability mass of all partial alignments $y_1^t$ up to time frame $t$ for the partial sequence $a_1^s$:\\
\scalebox{0.9}{\parbox{1.1\linewidth}{%
\begin{align*}
    q_\text{seq}(a_1^s|X,t) &=  \sum_{y_1^t: \mathcal{B}(y_1^t)=a_1^s}q_\text{seq}(y_1^t|X)\\
    &= \sum_{y_1^t: \mathcal{B}(y_1^t)=a_1^s} \prod_{\tau=1}^t q_\text{seq}(y_{\tau}|a_{s_{\tau-1}-k+1}^{s_{\tau-1}}, h_{\tau})
\end{align*}}}
where $q_\text{seq}(y_{\tau}|a_{s_{\tau-1}-k+1}^{s_{\tau-1}}, h_{\tau})$ is defined as:\\
\scalebox{0.9}{\parbox{1.1\linewidth}{%
\begin{align*}
    \begin{cases}
    P^\alpha_\text{RNNT}(\epsilon|a_{s_{\tau-1}-k+1}^{s_{\tau-1}},h_{\tau}), &y_{\tau}=\epsilon \\
    P^\alpha_\text{RNNT}(a|a_{s_{\tau-1}-k+1}^{s_{\tau-1}},h_{\tau}) \cdot P^\beta_\text{LM}(a|a_{s_{\tau-1}-k+1}^{s_{\tau-1}}), &y_{\tau} = a \in \mathcal{V}
    \end{cases}    
\end{align*}}}
Then \Cref{eq:mmiauxi} can be computed by DP recursion:\\
\scalebox{0.9}{\parbox{1.1\linewidth}{%
\begin{align*}
\begin{split}
    Q_\text{MMI}(t,u_1^k) &= Q_\text{MMI}(t-1,u_1^k) \cdot q_\text{seq}(\epsilon|u_1^k,h_t)\\
            +&  \sum_{u_0} Q_\text{MMI}(t-1, u_0^{k-1}) \cdot q_\text{seq}(u_k|u_0^{k-1},h_t)
\end{split}
\label{eq:DP_MMI}
\end{align*}}}
\vspace{-1mm}

With LM integration in training, the transducer model gathers external phoneme information to suppress unlikely sequences in the denominator. As discussed in \cite{zhou2022language}, CE-FS trained transducer models usually have a quite high blank probability. However, for LF-MMI, since the LM probability is smaller for longer label sequences, the model tends to assign large probabilities to long label sequences when minimizing the denominator. 
This leads to higher probabilities for labels, which mitigates the `dominant blank' issue.

\subsection{Segment-Level MBR}
To apply the above DP concept to MBR training in a LF manner, the biggest challenge is to design a cost function $\boldsymbol{R}$ feasible for the recombination scheme.
sMBR computes the cost locally per frame, which is compatible with LF training. However, in sMBR there is only one alignment regarded as the correct alignment, which is in contrast to the full-sum computation of $P_\text{RNNT}$. To allow small shifts of alignments, and make costs similar (at least locally) for different alignments corresponding to the same label sequence, we propose the LF segment-level MBR (LF-SegMBR), which computes costs according to the label of each segment generated from a target alignment $\hat{y}_1^T$.\\
\scalebox{0.95}{\parbox{1.05\linewidth}{%
\begin{equation}
    \mathcal{L}_\text{SegMBR} = \sum_{y_1^T} \frac{q_\text{seq}(y_1^T|X)}{\sum_{{y'}_1^T}q_\text{seq}({y'}_1^T|X)} \boldsymbol{R}(y_1^T, \hat{y}_1^T)
    \label{eq:SegMBR}
\end{equation}}}

The Viterbi alignment $\hat{y}_1^T$ can be generated from the baseline model for the training data, which also reveals the segment boundaries $\hat{t}_1^{\hat{S}}$ and the position sequence $\hat{s}_1^T$. 
The cost function $ \boldsymbol{R}= \boldsymbol{R}_1+ \boldsymbol{R}_2$ consists of two parts: the label-based cost function $ \boldsymbol{R}_1$ and the label emission penalty $ \boldsymbol{R}_2$. 
For $\boldsymbol{R}_1$, we map the blanks in $y_1^T$ to their previous labels by a mapping function $\mathcal{M}$. For instance, an alignment sequence $(a,\epsilon,\epsilon,b,\epsilon)$ is mapped to $(a,a,a,b,b)$. Besides, we introduce a smoothed cost function over a window $\hat{a}_{\hat{s}_t-L}^{\hat{s}_t+L}$ of length $2L+1$ to enable small shifts for alignments.\\
\scalebox{0.9}{\parbox{1.1\linewidth}{%
\begin{align}
    \notag \boldsymbol{R}_1(y_1^T, \hat{y}_1^T) &= \sum_{t=1}^T r(\mathcal{M}_t(y_1^T), \hat{a}_{\hat{s}_t-L}^{\hat{s}_t+L})\\
       r(a, \hat{a}_{-L}^L) &=  \left \{
     \begin{array}{cc}
        \underset{l: -L\le l \le L, \hat{a}_l=a}{\min} \frac{|l|}{L} \quad&,  a \in \hat{a}_{-L}^L  \\
        1  \quad&, \text{otherwise}
    \end{array}
    \right .
    \label{eq:smoothfun}
\end{align}}}
One problem of $\boldsymbol{R}_1$ is that emitting the correct label multiple times in one segment will not be penalized. For instance, if target label is $a$ and the partial alignment hypothesis for the segment is $y_{\hat{t}_{s-1}+1}^{\hat{t}_s}=(a,a,a)$, all the emissions of $a$ in this segment will be considered as correct in $R_1$. To penalize such emissions, we introduce a label emission penalty $\boldsymbol{R}_2$:\\
\scalebox{0.9}{\parbox{1.1\linewidth}{%
\begin{equation*}
    \boldsymbol{R}_2(y_1^T, \hat{y}_1^T) = \sum_{s=1}^{\hat{S}} f(\boldsymbol{N}(y_{\hat{t}_{s-1}+1}^{\hat{t}_s}))
\end{equation*}}}
where $\boldsymbol{N}(y_{\hat{t}_{s-1}+1}^{\hat{t}_s})$ is the number of emitted labels in the segment $s$ and $f$ is a penalty function. Here we choose $f(i)= c \cdot \max(i-1,0)$, which has a linear penalty with slope $c$ for the alignment sub-sequence emitting more than one label. 


For the time frames $t \in [\hat{t}_{s-1}+1, \hat{t}_s]$ in segment $s$, the auxiliary function $Q_s(t,i, u_1^k)$ for SegMBR is defined as:\\
\scalebox{0.95}{\parbox{1.00\linewidth}{%
\begin{equation}
    Q_s(t,i, u_1^k) = \sum_{y_1^t: \substack{\mathcal{B}(y_1^t)_{s'-k+1}^{s'}=u_1^k,\\ 
    N(y_{\hat{t}_{s-1}+1}^t)=i} }(q_\text{seq}(y_1^t|X), q_\text{seq}(y_1^t|X) \cdot \mathcal{R})
    \label{eq:smbrdef}
\end{equation}}}
where $i$ denotes the number of emissions in segment $s$ and $\mathcal{R}$ is the corresponding cost for the partial alignment $y_1^t$. \Cref{eq:smbrdef} can be calculated by DP with the expectation semiring \cite{eisner2001expectation}, we refer the reader to \cite{eisner2001expectation} for more details.


Besides the penalty for emissions, we also have a hard constraint that sequences with more than $I$ emissions in the segment are pruned out. At the end of each segment, the auxiliary functions are multiplied with the emission penalty and summed up over $i$ as the initialization for the next segment.\\
\scalebox{0.9}{\parbox{1.1\linewidth}{%
\begin{align}
    Q_{s+1} (t_{s}, 0 ,u_1^k) = Q_{s} (t_{s},u_1^k) 
                              = \sum_{i=0}^I Q_{s} (t_{s}, i ,u_1^k) \otimes (1,f(i))
\end{align}}}
the final auxiliary function $Q_{\hat{S}} (T)$ then computes the numerator and denominator for \Cref{eq:SegMBR}.\\
\scalebox{0.9}{\parbox{1.1\linewidth}{%
\begin{align}
\begin{split}
    Q_{\hat{S}} (T) &= \sum_{u_1^k} Q_{\hat{S}} (T,u_1^k) =( Q^1_{\hat{S}} (T), Q^2_{\hat{S}} (T))\\
    &=  (\sum_{{y}_1^T}q_\text{seq}({y}_1^T|X),\sum_{y_1^T} q_\text{seq}(y_1^T|X) \boldsymbol{R}(y_1^T, \hat{y}_1^T))
\end{split}
\label{eq:smbrfinal}
\end{align}}}

\subsection{Label-based MBR}
In this section, we consider the cost function on the output label sequence level, i.e. given a label sequence $a_1^S$, each alignment $y_1^T \in \mathcal{B}^{-1}(a_1^S)$ has exactly the same risk, which is consistent with the computation of label sequence probabilities. The objective of LF-MBR is formulated as:\\
\scalebox{0.9}{\parbox{1.10\linewidth}{%
\begin{equation}
    \mathcal{L}_\text{LF-MBR} = \sum_{S,a_1^S} \frac{q_\text{seq}(a_1^S|X)}{\sum_{S',{a'}_1^{S'}}q_\text{seq}({a'}_1^{S'}|X)}R(a_1^S, \hat{a}_1^{\hat{S}})
    \label{eq:hamming}
\end{equation}}}
In \cite{Guo2020EfficientMW, zhou2022efficient}, the word-level edit distance is applied as the cost function for N-best MBR, which is consistent with the metric for the performance measurement. However, a word-level Levenshtein alignment between reference and hypothesis is needed, which cannot be obtained locally, and thus it is not feasible for LF methods. Meanwhile, Hamming distance, which effectively compares labels in reference and hypothesis per position, can be computed locally and suits the requirement for recombination in DP. To avoid the alignment problem, we use phoneme-level Hamming distance with a smoothing window to be the risk function for LF-MBR. The cost function is defined as:\\
\scalebox{0.9}{\parbox{1.2\linewidth}{%
\begin{equation*}
    R(a_1^S, \hat{a}_1^{\hat{S}}) = \sum_{s=1}^{S_\text{pad}} r(a_s, \hat{a}_{s-L}^{s+L})
\end{equation*}}}
Here $r$ is the same smoothed cost function defined in \Cref{eq:smoothfun}. The cost is computed per position, and both label sequences are padded at the end to the same length $S_\text{pad}= \max\{S,\hat{S}\}$ in order to compute the risk for each position. 
Similar to \Cref{eq:SegMBR}, \Cref{eq:hamming} can be computed by DP. The auxiliary function $Q_\text{MBR}(t,s, u_1^k)$ is defined and computed similarly to \Cref{eq:smbrdef}, with $s$ denoting the position in the output label sequence. 

We assume that sequences with low probabilities are quite different from the target sequence, which might bring in harmful cost information because of the difference between Hamming and Levenshtein distance. Therefore, we prune out sequences with low probabilities at each time frame. Similar to \Cref{eq:smbrfinal}, $Q^1_\text{MBR}(t,s,u_1^k)$ denotes the probability mass for partial sequences. The prune factor is computed by $\mu_{t,s} = \max_{u_1^k} Q^1_\text{MBR}(t,s,u_1^k)$. The sub-sequences with $Q^1_\text{MBR}(t,s,u_1^k) <\mu_{t,s}^\gamma$ are pruned out where $\gamma > 1$ is a scale.

Due to the memory constraint, we use a Viterbi alignment to obtain the target length $\hat{s}_t$ of the label sequence and generate a length window at each time frame $t$. When computing $Q_\text{MBR}$, only sequences with length in the window are kept, other sequences that are too long or too short are pruned out.

\section{Experiments}
\label{sec:experiments}

\subsection{Experimental Setup}

We conduct our experiments on 960h Librispeech (LBS) \cite{panayotov2015librispeech}. The architecture of the transducer model follows \cite{zhou2022efficient}. We use 12 conformer \cite{Gulati2020ConformerCT} layers as encoder and 2 feed-forward layers as prediction network. We employ gammatone features \cite{schluter2007gammatone} with 50 dims as the input.
The bi-gram phoneme LM used in training is trained only on transcripts of LBS, with the same architecture as the prediction network, followed by a softmax.

In training, we follow the pipeline proposed in \cite{zhou2022efficient}. We use LF objectives in two ways: the first one is to fine-tune the CE-FS trained model with LF objectives, and the second one is to only do Viterbi training as initialization, and then directly train the model with LF-MMI. All hyperparameters are tuned on dev set. For all three LF objectives, we choose $\alpha=1.2$ and $\beta=0.3$. For LF-SegMBR, we use $I=3$, $L=3$ and $c=0.3$. For LF-MBR, the size of the pruning window is 4 and $\gamma=1.1$. LF-MBR and LF-Seg MBR are integrated with LF-MMI with 0.2 as the scale during training. We use a 4-gram word level LM to generate N-best list with $N=4$. We mainly compare our methods to the N-best-list based MBR. For decoding, we apply 1-pass SF decoding with a word-level transformer (trafo) LM following the setup in \cite{Irie2019LanguageMW}. 


\begin{table}[t]
\caption{Comparison of different criteria with external LM integration on LBS dev-other}
\centering
\scalebox{0.85}{\parbox{1\linewidth}{%
\begin{tabular}{|l|cc|cccc|}
\hline
Objective   & \multicolumn{2}{c|}{LM}      & \multicolumn{4}{c|}{Dev-other [\%]}                                                             \\ \cline{2-7} 
            & \multicolumn{1}{c|}{$\lambda_1$} & $\lambda_2$ & \multicolumn{1}{c|}{Sub} & \multicolumn{1}{c|}{Del} & \multicolumn{1}{c|}{Ins} & WER \\ \hline
CE-FS         & \multicolumn{1}{c|}{1.0}   &  0.2  & \multicolumn{1}{c|}{3.1} & \multicolumn{1}{c|}{0.4} & \multicolumn{1}{c|}{0.4} & 3.9 \\ \cline{2-7} 
+N-best MBR & \multicolumn{1}{c|}{1.3}   &  0.0  & \multicolumn{1}{c|}{3.0} & \multicolumn{1}{c|}{0.4} & \multicolumn{1}{c|}{0.4} & 3.7 \\ \cline{2-7} 
+LF-MMI     & \multicolumn{1}{c|}{1.0}   &  0.1  & \multicolumn{1}{c|}{3.1}    & \multicolumn{1}{c|}{0.3} & \multicolumn{1}{c|}{0.4}    & 3.7 \\ \cline{2-7} 
+LF-SegMBR  & \multicolumn{1}{c|}{1.3}   &  0.0  & \multicolumn{1}{c|}{3.1} & \multicolumn{1}{c|}{0.3} & \multicolumn{1}{c|}{0.4} & 3.8 \\ \cline{2-7} 
+LF-MBR     & \multicolumn{1}{c|}{1.2}   &  0.2  & \multicolumn{1}{c|}{3.1} & \multicolumn{1}{c|}{0.3} & \multicolumn{1}{c|}{0.4} & 3.8 \\ \hline
\end{tabular}}}
\label{tab:lmintegrate}
\vspace{-2.5mm}
\end{table}

\begin{table}[t]
\caption{Training speed of N-best vs LF-based methods (on one single 1080Ti GPU)}
\centering
\scalebox{0.85}{\parbox{1\linewidth}{%
\begin{tabular}{|c|ccc|}
\hline
\multirow{2}{*}{Objective} & \multicolumn{3}{c|}{\begin{tabular}[c]{@{}c@{}}Training speed\\ (hours/epoch)\end{tabular}} \\ \cline{2-4} 
                           & \multicolumn{1}{c|}{Training}        & \multicolumn{1}{c|}{Decoding}        & Total        \\ \hline
N-best MBR                 & \multicolumn{1}{c|}{30}                & \multicolumn{1}{c|}{111}                &    141          \\ \hline
LF-MMI                     & \multicolumn{1}{c|}{43}                & \multicolumn{1}{c|}{-}                &      43        \\ \hline
LF-SegMBR                  & \multicolumn{1}{c|}{80}                & \multicolumn{1}{c|}{-}                &       80       \\ \hline
LF-MBR                     & \multicolumn{1}{c|}{75}                & \multicolumn{1}{c|}{-}                &       75       \\ \hline
\end{tabular}}}
\label{tab:speed}
\vspace{-5mm}
\end{table}

\subsection{Sequence Training for CE-FS trained Model}
Table \ref{tab:lmintegrate} shows the results of different criteria with LM integration. In our experiments, N-best MBR and LF-MMI need to use 10\% of total training data for fine-tuning to have the best performance, while LF-SegMBR and LF-MBR need 5\%. For LF-MMI, as discussed above, the model tends to output longer sequences and has fewer Del errors compared to CE-FS. For LF-SegMBR, according to the design of the risk function, the model is penalized for outputting a blank label with a wrong context and encouraged to output the correct label at any frame within the segment, which eventually assigns large probabilities to labels and allows a large LM scale $\lambda_1$ even without ILM correction ($\lambda_2=0)$.
For LF-MBR, although Hamming distance is not a good approximation of Levenshtein distance, it still brings useful cost information for the sequence, which helps close the gap between training and evaluation and improve performance. Since the CE-FS trained model is already well-tuned on dev-other, the performance gains from sequence training are not so large: 5\% relative improvement for N-best MBR and LF-MMI, and 2.5\% for LF-SegMBR and LF-MBR.

Table \ref{tab:speed} shows training efficiency for different sequence training criteria. Although the pure training speed of LF methods is slower than N-best MBR, due to the extra computation for all possible sequences, the total training time for LF methods is much less than N-best MBR since there is no decoding step needed. LF-SegMBR and LF-MBR are slower than LF-MMI because of the extra computation for the expectation ring. Overall, LF-MMI training gives relative training time speedup of 70\% compared to N-Best MBR, while for LF-SegMBR and LF-MBR gain speedup of over 40\%. 
\begin{table}[t]
\caption{Overall WER [\%] results on LBS (* means some sequences are pruned out)}
\setlength{\tabcolsep}{0.2em}
\scalebox{0.75}{\parbox{1.1\linewidth}{%
\begin{tabular}{|l|c|c|cccc|}
\hline
\multirow{3}{*}{Objective} & \multirow{3}{*}{\begin{tabular}[c]{@{}c@{}}Hypotheses \\ space \end{tabular}} & \multirow{3}{*}{\begin{tabular}[c]{@{}c@{}}Cost \\ function\end{tabular}} & \multicolumn{4}{c|}{WER {[}\%{]}}                                                            \\ \cline{4-7} 
                           &                                   &                                & \multicolumn{2}{c|}{dev}                                & \multicolumn{2}{c|}{test}          \\ \cline{4-7} 
                           &                                   &                                & \multicolumn{1}{c|}{clean} & \multicolumn{1}{c|}{other} & \multicolumn{1}{c|}{clean} & other \\ \hline
CE-FS                        &             -                     &             -                  & \multicolumn{1}{c|}{1.8}   & \multicolumn{1}{c|}{3.9}   & \multicolumn{1}{c|}{2.1}   & 4.6   \\ \cline{2-7}
+N-best MBR                &             N-best           &        {\begin{tabular}[c]{@{}c@{}} phoneme-level \\ edit distance\end{tabular}}           & \multicolumn{1}{c|}{1.7}   & \multicolumn{1}{c|}{3.7}   & \multicolumn{1}{c|}{2.1}   & 4.1   \\ \cline{2-7}
+LF-MMI                    &             all seq        &             -                  & \multicolumn{1}{c|}{1.7}   & \multicolumn{1}{c|}{3.7}   & \multicolumn{1}{c|}{2.1}   & 4.3   \\ \cline{2-7}
+LF-SegMBR                 &             all seq*        &     {\begin{tabular}[c]{@{}c@{}}    phoneme\\ segment-level \end{tabular}}   & \multicolumn{1}{c|}{1.7}   & \multicolumn{1}{c|}{3.8}   & \multicolumn{1}{c|}{2.1}   & 4.3   \\ \cline{2-7}
+LF-MBR                    &             all seq*        &    {\begin{tabular}[c]{@{}c@{}} phoneme-level \\ Hamming distance\end{tabular}}  & \multicolumn{1}{c|}{1.7}   & \multicolumn{1}{c|}{3.8}   & \multicolumn{1}{c|}{2.1}   & 4.3   \\ \hline
\end{tabular}}}
\label{tab:compare_LBS}
\vspace{-3mm}
\end{table}

Table \ref{tab:compare_LBS} shows the overall performance for different criteria on LBS. LF methods obtain about 7\% relative improvements on test-other compared to CE-FS baseline. Compared to N-best MBR, LF methods are slightly worse on test-other, but comparable on the other three datasets.

\subsection{Sequence Training for Viterbi Initialized Model}
\begin{table}[t]
\centering 
 \caption{Comparison of CE-FS and LF-MMI training (initialized with the same Viterbi trained model) on LBS}
\scalebox{0.85}{\parbox{1\linewidth}{%
\begin{tabular}{|c|c|cccc|}
\hline
\multirow{3}{*}{Objective} & \multirow{3}{*}{epochs} & \multicolumn{4}{c|}{WER [\%]}                                                                     \\ \cline{3-6} 
                           &                         & \multicolumn{2}{c|}{dev}                                & \multicolumn{2}{c|}{test}          \\ \cline{3-6} 
                           &                         & \multicolumn{1}{c|}{clean} & \multicolumn{1}{c|}{other} & \multicolumn{1}{c|}{clean} & other \\ \hline
CE-FS                         & 15                      & \multicolumn{1}{c|}{1.8}      & \multicolumn{1}{c|}{3.9}   & \multicolumn{1}{c|}{2.1}      &   4.6    \\ \hline
LF-MMI                     & 9.6                     & \multicolumn{1}{c|}{1.8}      & \multicolumn{1}{c|}{3.8}   & \multicolumn{1}{c|}{2.1}      &   4.5    \\ \hline
\end{tabular}}}
\label{tab:viterbi}
\vspace{-5mm}
\end{table}
LF-MMI can also be used to replace CE-FS for from-scratch training or fine-tuning the Viterbi-trained model. Due to hardware and time constraint, we investigate the effect of LF-MMI for fine-tuning the Viterbi-trained model. Table \ref{tab:viterbi} shows that with LF-MMI, the model gains slightly better performance with fewer training epochs.

\vspace{-0.5mm}
\section{Conclusion}
\vspace{-0.5mm}
In this paper, we propose three lattice-free (LF) methods (MMI, Segment-level MBR, and label-based MBR) applied directly to the final posterior outputs of neural transducers with limited context dependency. We show how the objectives are calculated by dynamic programming in detail. For MBR-based objectives, we design two cost functions that are suitable for LF computation.
Compared to N-best-list based methods, these LF methods eliminate the need for decoding in training, which leads to more efficient training. 
Experiments on Librispeech show that LF methods can obtain 40\% - 70\% relative training speedup with a slight degradation in performance compared to N-best-list MBR. 
Furthermore, we show that LF-MMI can be used to replace standard cross-entropy training of transducer model, where the model can converge with fewer epochs and obtain a slightly better performance.
{
\footnotesize
\section{Acknowledgments}
}
\begin{spacing}{0.1}
{
\scriptsize
\selectfont

This work was partially supported by the project HYKIST funded by the German Federal Ministry of Health on the basis of a decision of the German Federal Parliament (Bundestag) under funding ID ZMVI1-2520DAT04A.
This work was partially supported by NeuroSys which, as part of the
initiative “Clusters4Future”, is funded by the Federal Ministry of
Education and Research BMBF (03ZU1106DA).
This work was partially supported by a Google Focused Award. The work
reflects only the authors' views and none of the funding parties is
responsible for any use that may be made of the information it contains.
}

\end{spacing}

\let\normalsize\small\normalsize
\let\OLDthebibliography\thebibliography
\renewcommand\thebibliography[1]{
        \OLDthebibliography{#1}
        \setlength{\parskip}{-0.3pt}
        \setlength{\itemsep}{0.5pt plus 0.05ex}
}
\bibliographystyle{IEEEbib}
{\footnotesize
\bibliography{strings,refs}}

\end{document}